\documentclass[twocolumn,pra,amssymb,showpacs]{revtex4-1}
\usepackage{graphicx}
\usepackage{amsmath}
\usepackage{subfigure}

\newcommand{\be}{\begin{equation}}
\newcommand{\ee}{\end{equation}}
\newcommand{\ben}{\begin{eqnarray}}
\newcommand{\een}{\end{eqnarray}}

\newcommand{\nd}{\noindent}

\begin{document}

\title{Generalized Thomson problem in arbitrary dimensions and non-euclidean geometries}
\author{J. Batle$^1$}
\email{E-mail address: jbv276@uib.es}
\affiliation{ $^1$Departament de F\'{\i}sica, Universitat de les Illes Balears,
 07122 Palma de Mallorca, Balearic Islands, Europe  \\\\}

\date{\today}

\begin{abstract}

\nd Systems of identical particles with equal charge are studied under a special type of confinement. These classical particles are free to move 
inside some convex region $S$ and on the boundary of it $\Omega$ (the $S^{d-1}-$sphere, in our case). We shall show how particles arrange themselves 
under the sole action of the Coulomb repulsion in many dimensions in the usual Euclidean space, therefore generalizing the so called {\it Thomson problem} 
to many dimensions. Also, we explore how the problem varies when non-Euclidean geometries are considered. We shall see that optimal configurations 
in all cases possess a high degree of symmetry, regardless of the concomitant dimension or geometry.

\end{abstract}


\maketitle

\section{Introduction}

Since Wigner's prediction \cite{Wig} of the crystallization of electrons, Wigner crystals have attracted the attention of a 
wide variety of research communities.  At first the attention of research was focused on electron crystals in solid 
state systems, a setting for which the
concept of Wigner crystallization had originally been conceived. However, their realization has proven to be
very challenging. In fact, it took almost half a century before the first creation of a Wigner crystal on the surface of
superfluid Helium could be realized \cite{Grimes}, and a few more years until Wigner crystals were achieved in 
GaAs/GaAlAs quantum wells \cite{Andrei}. Only recently a 1D Wigner crystal was observed \cite{Deshpande} in carbon nanotubes. 

Posterior research on the original Wigner problem triggered a considerable body of experimental and theoretical work on the properties of
ionic Coulomb crystals, and that is where we encounter the classical {\it Thomson problem} \cite{tom}. 
The original goal of the Thomson problem is the following: given $N$ charges confined to the surface of a sphere, what is the arrangement of charges which minimizes theÊtotal electrostatic energy? In essence, the Thomson problem is concerned with finding the minimal energy ground state of a cluster of charges in an arbitrary 
geometries and nature of confinements, not only on the $S^2-$sphere.
 
Also, the Thomson problem is widely regarded as one of the most important unsolved packing problem in mathematics. It is also important for two reasons. 
On the one hand, it plays a central role in the field of strongly correlated Coulomb systems such as quantum dots, dusty plasmas, and colloidal crystals. On the other hand, the Thomson problem yields fundamental insights into the interplay of geometry and topology in ordered systems. Specifically, systems with planar geometry have been studied previously \cite{loz,bol,pet1,pet2}.

Since the original case was initially intended for the $S^1$ (2D) and $S^2$ (3D) spheres in the usual Euclidean 
space, it is the aim of the present work to generalize the problem to those systems which live in 

\begin{itemize}
\item higher dimensional $S^{d-1}$-spheres, $d$ being the dimension of the concomitant Euclidean space, 
\item and systems where the metric space is changed so that it is no longer Euclidean (Elliptic $\mathbb{E}^d$ 
or Hyperbolic $\mathbb{H}^d$).
\end{itemize}

\noindent The study of non-Euclidean geometries and the Thomson problem bears great significance as far as 
geometry and physics are concerned. The way optimal energies $E_N$, as we shall see, behave differently for 
distances are measured using distinct metrics. As expected, minimal configurations and regular bodies will be 
intimately related. There is a physical reason for that since regular bodies are such that the sum of their respective vector 
positions $\sum_i {\bf r}_i$ is zero, which implies that there is no net dipole moment.

Let us briefly discuss the numerical methods for obtaining the exact energies and configurations for 
any confinement throughout this work. When working in the definite plane or space according to some metric, 
we will have $k$ degrees of freedom per charged particle. Thus, the total number of variables will be $kN$. 
A minimization will take place for the whole set of parameters in every given configuration of the particles, 
finding the optimal Coulombian energy $E_N^{*}$. 

The Thomson problem is certainly the kind of example of an NP hard problem and so progress in this area has only 
been possible thanks to the use of computational techniques. In our case, we have performed a two-fold search employing 
i) an amoeba optimization procedure, where the optimal value 
is obtained at the risk of falling into a local minimum and ii) the so called simulated annealing \cite{kirkpatrick83} 
well-known search method, a Monte Carlo method, inspired by the cooling processes of molten metals. 
The advantage of this duplicity of computations 
is that we can be absolutely confident about the final result reached. Indeed, the second recipe contains a mechanism 
that allows a local search that eventually can escape from local optima.

The purpose of this paper is to provide a semi-analytical approach to describe 
the ground state properties of charged particles in different geometries. Specially in the 
case of the $S^{d-1}$-sphere, we shall consider the minimal energies and concomitant configurations or 
arrangements of charged particles in different dimensions and, eventually, reach the 
limit for $d \rightarrow \infty$.. We consider particles
interacting by means of the Coulomb interaction at zero temperature. Finally, some conclusions are 
drawn in the last Section.

\section{The Thomson problem in arbitrary dimensions}

\subsection{One dimension}

As explained previously, the only interaction between particles is electrostatic in nature. This implies that 
all particles interact with each other until an energetic equilibrium is reached, which is the one we are interested in. 
Suppose that we want to study the system composed by charged particles along a line segment between [-$R_0$,$R_0$] 
(the radius $R_0$ will be 1 from now on). Due to symmetry reasons, the system is symmetric with respect to the center. 
For $N$ even, no particle lies at the origin, whereas for $N$ odd there is always one charge. Even though the physical 
system is quite simple, the optimal configuration of the charges for a minimum energy $E_N$ is not analytical. 

However, we can obtain an excellent upper bound by considering that, for large $N$, particles more or less arrange themselves 
in equally spaced divisions of the line segment containing them. Thus, defining the linear charge density $\lambda=\frac{L}{N-1}$ 
($L=2$ in our case), we have

\begin{equation} \label{1D}
E_N\,\approx \,\sum_{i<j} \frac{1}{\epsilon}\frac{1}{(j-i)\lambda}\,=\,\frac{N-1}{2\epsilon_r} \sum_{i<j} \frac{1}{j-i}.
\end{equation}

\noindent where $\epsilon_r=\epsilon/\epsilon_0$ is the relative dielectric constant of the medium. 
We shall use units so that $e^2/4\pi\epsilon_0=1$ from now on. The sum is performed between distinct pairs, so that it is equal to $1+(H(N-1)-1)N$, 
where $H(N-1)$ is the sum of the harmonic series up to $N-1$. This relation ca be easily proved by induction. From $H(N-1)=\ln(N-1)+\gamma + O(1/N)$, 
$\gamma$ being the Euler constant, we derive the following bound for large $N$

\begin{equation} \label{1D2}
E_N\,<\,\frac{N}{2\epsilon_r}\, N\,\ln N,
\end{equation}

In the usual definition of the electrostatic energy of a discrete distribution of charged particles, we have $E=\frac{Q^2}{2C}$, $Q$ 
being the total charge and $C$ the capacity of the system. Usually, the capacity is either constant of geometry dependent. In our 
case, if will be definitely determined by both the geometry, the type of confinement and the number of particles $N$. 

\noindent Thus, we shall have $E_N=\frac{N^2}{2C_N}$. From this observation, it will prove convenient to consider the 
quantity $E_N/N^2$ rather than $E_N$. Therefore, we can conclude from (\ref{1D2}) that $E_N/N^2$ diverges logarithmically as the total number 
of particles $N$ tends to infinity. The actual dependence can either be that of a power law $N^{\alpha}$, with $0 <\alpha<1$, or corresponding to 
$\log N \cdot f(M)$, with $f(N)$ being a correction term. Knowing the form of the divergence is interesting since it provides us with a tool in order to compare 
different confinement geometries. 

\begin{figure}[htbp]
\begin{center}
\includegraphics[width=8.8cm]{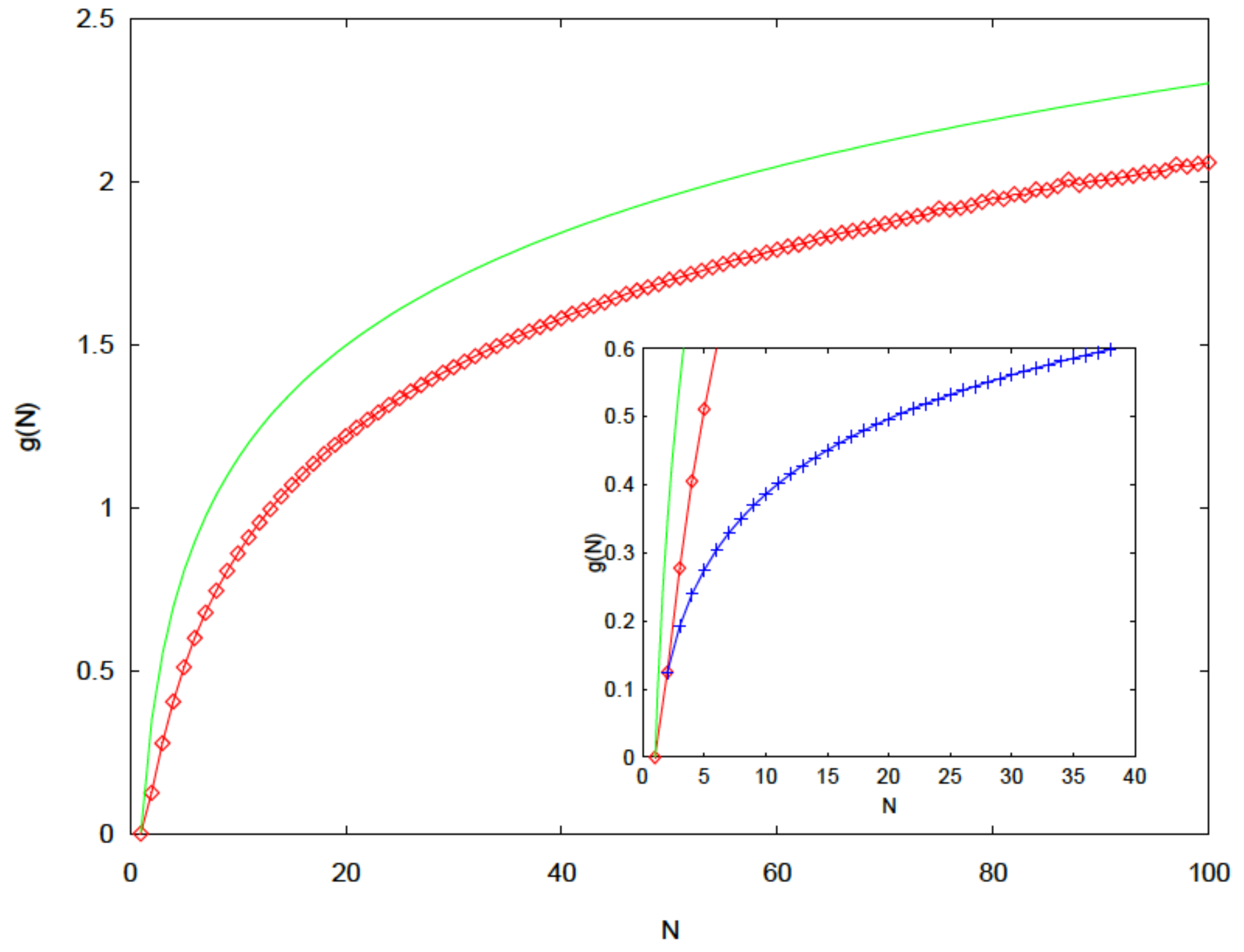}
\caption{(Color online) The red curve represents the exact numerical results for $E_N/N^2$ in the 1D system of charges. The upper (green) curve is the analytic bound 
encountered. The inset compares the radical change when considering the 2D circular case (blue line with crosses). See text for details.}
\label{FigA}
\end{center}
\end{figure}

Now, what is the exact functional dependency of $E_N/N^2$? In order to answer this question we have to resort to numerical computations. Given a particular 
number of particles $N$, we must found the optimal distribution of particles that ensues a minimal value for $E_N/N^2$. By recourse to a simulated annealing 
Monte Carlo computation, we have computed these equilibrium values (notice that we have $N$ total degrees of freedom). 
These ones appear depicted in Fig. (\ref{FigA}). The point-like red curve is the exact numerical result for optimal $E_N/N^2$, while the upper green line represents 
the concomitant upper bound $\frac{1}{2} \ln N$. We can appreciate that the asymptotic behavior is reached at some point between $N=20$ and $N=40$. Therefore, 
at this point we may have $E_N/N^2 \approx A \log \sqrt N + B$, where $A,B$ are some constants. 
In the inset of the same Fig. (\ref{FigA}), the numerical results for the 2D circular case (blue crosses) are depicted for comparison. Notice the abrupt change in the 
corresponding slopes.

It is remarkable to mention that the one dimensional case of the Thomson problem possesses a very interesting connection with 
orthogonal polynomials in one variable. The x-zeros of the Jacobi polynomial $P_n^{\alpha,\beta}(\cos\theta)$ may be thought of as the 
positions of equilibrium of $n$ unit electrical charges in the interval (-1,1), with logarithmic potential $-\ln|x_i-x_j|$, generated 
by charges $(\alpha+1)/2$ and $(\beta+1)/2$ placed at 1 and -1 (in our case $\alpha=\beta=1$). This interpretation is originally 
due to Stieltjes (see \cite{Szego} for a proof, and references therein). Also, in the limit of very large number of charges, the corresponding 
density of charges/zeros tends to (see, for instance, Ref. \cite{Dehesa}) the so-called {\it arc-sin} density $\rho(x)=\frac{1}{\pi\sqrt{1-x^2}}$. 

The actual distribution of charges is not of the previous form due to the Coulombian nature of the interaction: it is very uniform, as opposed to the arc-sin 
distribution, but also peaks at the extremes. Actually, it is this uniformity in the position of the charges that proves (\ref{1D}) to be correct asymptotically. 

What is certainly surprising is that, as shown in Ref. \cite{Ismail}, we obtain the same large $N$ asymptotics for $E_N$, given by (\ref{1D2}), regardless of the 
repulsive force. It is surprising that any repulsive potential can give rise to the same energy asymptotics, having an entirely mathematical approach to the 
electrostatic problem based on orthogonal polynomials.

\subsection{Two dimensions}

At first sight, the segment system does not bear much resemblance with the circular case, which we shall study here. 
Recall that, in mathematics, two sets A and B 
are topologically equivalent if there is a one-to-one correspondence 
between them which is continuous both ways. Thus, a circle is not (topologically) equivalent to a line segment, 
because you need to glue the two endpoints of the segment together in order to make a circle. However, if we regard 
the circle as a segment with periodic boundary conditions, we ought to obtain similar results when computing, say, the total 
Coulombian energy $E_N$ of the system. And that is indeed the case.

Let us consider a circle of unit radius, where all charges are allowed to stay at $R=1$. Again, we shall perform a Monte Carlo 
minimization over the $N$ total degrees of freedom of the system.

All particles lie on $r=R$. Under this assumption, the total energy is easily obtained since all particles are equally spaced 
in their equilibrium value. The minimal energy of the system is

\begin{equation} \label{EsinN}
E_N=\frac{1}{\epsilon_r} \sum_{i<j}\frac{1} {r_{ij}}=\frac{1}{2R\epsilon_r} \sum_{i<j}\frac{1} {\sin\frac{\pi}{N}(|i-j|)}.
\end{equation}

\noindent The previous expression can be approached analytically in the regime $N\rightarrow \infty$. In that case, 
the sinus term can be replaced by its argument, and thus obtaining 

\begin{equation} 
E^{'}_N \sim \frac{N}{2\pi R\epsilon_r} \sum_{i<j}\frac{1} {|i-j|}= \frac{N}{2\pi R\epsilon_r} \bigg( 1+(H(N-1)-1)N \bigg),
\end{equation}

\noindent where $H(N-1)$ is again the sum of the harmonic series up to $N-1$. The first finite-size correction term to (\ref{EsinN}) 
is $\frac{2\pi}{3!N}\sum_{i=1}^{[\frac{N-1}{2}]} i\,(N-i)$, which arises when taking into account the second term in the series expansion 
of $\sin \alpha$. In view of the nature of this correction term, it is unlikely to devise a clear analytical $N-$dependence for $E_N$.
0
From $H(N-1)=\ln(N-1)+\gamma + O(1/N)$, 
$\gamma$ being the Euler constant, we can easily derive the following bound:

\begin{equation} 
\frac{N^2}{2\pi R\epsilon_r} \ln N < E_N = \frac{1}{2R\epsilon_r} \sum_{i<j}\frac{1} {\sin\frac{\pi}{N}(|i-j|)}.
\end{equation}

The corresponding exact curve is depicted in the inset of Fig. (\ref{FigA}). Although the circular case differs quite abruptly from 
the linear case, both have the same asymptotic dependence. Thus, there is no great different between these two confinement 
geometries after all. 

\subsubsection{Circular hard-wall confinement}

When the previous system increases the number of particles beyond $N=11$, the remaining particles are forced to move inwards, something that does not 
occur in 3D. The circular 
case of $N$ identical charges that interact electrostatically under a hard-wall confinement has been extensively studied in the literature 
\cite{pet1,pet2}. The energies for the configurations of minimal energy are known to a very high precision, even reaching hundreds and thousands 
of particles. Our concern here is to review the Thomson problem by a new approach that makes use of the rotational symmetry of the system. 

Since the pattern displayed by particles for $N$ up to 11 is that of regular polygons, we may wonder whether that may hold for bigger $N$. 
It is seen that for cases between $N=12$ and $N=16$, it is energetically favourable to have one particle at the center. However, starting from 
$N=17$, new rings start to form. In view of these facts, we shall consider the system of total charges divided into compact, regular rings of 
charges. The outer ring coincides with the boundary, where new polygons are allowed to form inside. This approach has to be seen as an approximation, 
for it is quite unclear that all particles could arrange themselves forming exactly regular polygons. 

Let us consider the cases (from 17 to 29) whit a single interior ring. Our method has to be able to approximate the position of the inner radios and the 
total energy. Thus, the total energy functional form will be that of (\ref{EsinN}) for each shell plus the contribution arising from the interaction 
between shells. The relative orientation of the interior ring is given by an angle $\alpha$ and the radios $R$. Now the problems consists in 
optimizing the total energy $E_N$ with respect to $(\alpha,R)$. The outcome is given by $\alpha=0$ and some radios $R^{*}$. When comparing the 
results performed using this ``shell'' approximation with the ones obtained numerically, we obtain an excellent agreement with respect to the 
total energies (ranging from $2\%$ to $<0.5\%$). That is, the agreement improves monotonically as $N$ increases. The number of particles in each shell 
coincides perfectly with numerical computations because the optimal configuration is obtained by choosing the minimum minimorum among all individual 
shells for a given $N$.

For very large $N$, the approach to the the problem by using ring and interring interaction can be done in principle. Let us suppose that our systems is composed by a series of concentric rings, such that $N_1+N_2+N_3+..+N_n=N$, each one at different $r_i$, where we have $n$ rings each one containing $N_i$ particles. 
If we regard the total energy $E_N \approx \sum_i^n \frac{N_i^2}{2\pi R\epsilon_r} \ln N_i +\sum_{i<j}^{n(n-1)/2} E_{ij}$, with the interring energy $E_{ij}$ being of the form  $\frac{2N_i N_j}{\pi |r_i-r_j|} K\big(-\frac{4r_i r_j}{(r_i-r_j)^2}\big)$ ($K(\cdot)$ is the complete elliptic integral of the first kind), we clearly see that no clear 
form is obtained. Therefore, the approach becomes rather intractable in the asymptotic behavior. 

However, a proper approach for large $N$ is more direct in the continuous limit. The sums over energies can be approximated by integrals over the disc $r \le R$

\begin{equation} \label{2dint}
E=\frac{1}{2}\int\!\,d^2{r}\int\!\,d^2{r}\frac{\rho{({\bf r})}\rho{({\bf r'})}}{|{\bf r}-{\bf r'}|}.
\end{equation}
\noindent The continuum approximation treats the density $\rho({\bf r})$ as a smooth function of the radios $r$. The problem of optimizing $E$ reduces to a variational one with respect to $\rho({\bf r})$, subject to the constraint $N=\int\!\,d^2{r}\rho{({\bf r})}$.

Introducing the Lagrange multiplier $\lambda$, we make the functional derivative stationary provided that
\begin{equation}
\lambda=\int\!\,d^2{r}\frac{\rho{({\bf r})}}{|{\bf r}-{\bf r'}|}.
\end{equation}

To solve this integral equation it is convenient to write integral (\ref{2dint}) explicitly as

\begin{eqnarray}
\lambda&=&\int^{R}_{0}\!\,\rho(r)rdr
\int_{0}^{2\pi}\!\,\frac{d\theta}{\sqrt{r^{2} +{r'}^{2} -2rr'\cos \theta} }
\nonumber 
\\
&=&
\int^{R}_{0}\!\,dr\frac{4r\rho(r)}{r+r'}K\left(\frac{2\sqrt{rr'}}{r+r'}\right).
\nonumber
\end{eqnarray}

This Fredholm integral equation of the first kind possesses the solution (after obtaining $\lambda$ from normalization) 

\begin{equation}
\rho(r)=\frac{N}{2\pi R^2}\frac{1}
{\sqrt{ 1 - \left(\frac{r}{R}\right)^{2} } },
\end{equation}

\noindent or, equivalently, 

\begin{equation}
N(r)=N\left(1-\sqrt{1-\left(\frac{r}{R}\right)^2}
\right).
\end{equation}

\noindent We clearly see that the density peaks at the perimeter, that is, charges tend to accumulate on the boundary.

Now, since we are interested in finding the asymptotic behavior for $E_N$, we need to compute the `electrostatic energy' 

\begin{equation}
\frac{1}{2}\int\!\,\rho(r)d^2{ r}\int\!\,\rho(r')d^2{r'}
\frac{1}{|{\bf r} - {\bf r}'|}=\frac{\pi}{4}\frac{N^2}{R}.
\end{equation}

\noindent Since this is the continuum limit approximation, additional terms have to be subtracted because the charges
are discrete. However, the leading order is of $O(N^2)$, which implies that, from $E_N=\frac{N^2}{2C_N}$, the capacity of the system 
when $N \rightarrow \infty$ is does not go to zero. This fact certainly constitutes a paradoxical result in the study of infinitely many identical charges confined 
in an infinite hard-wall circular potential.

\subsubsection{Circumscribed and inscribed charged polygons}

\begin{figure}[htbp]
\begin{center}
\includegraphics[width=5.1cm]{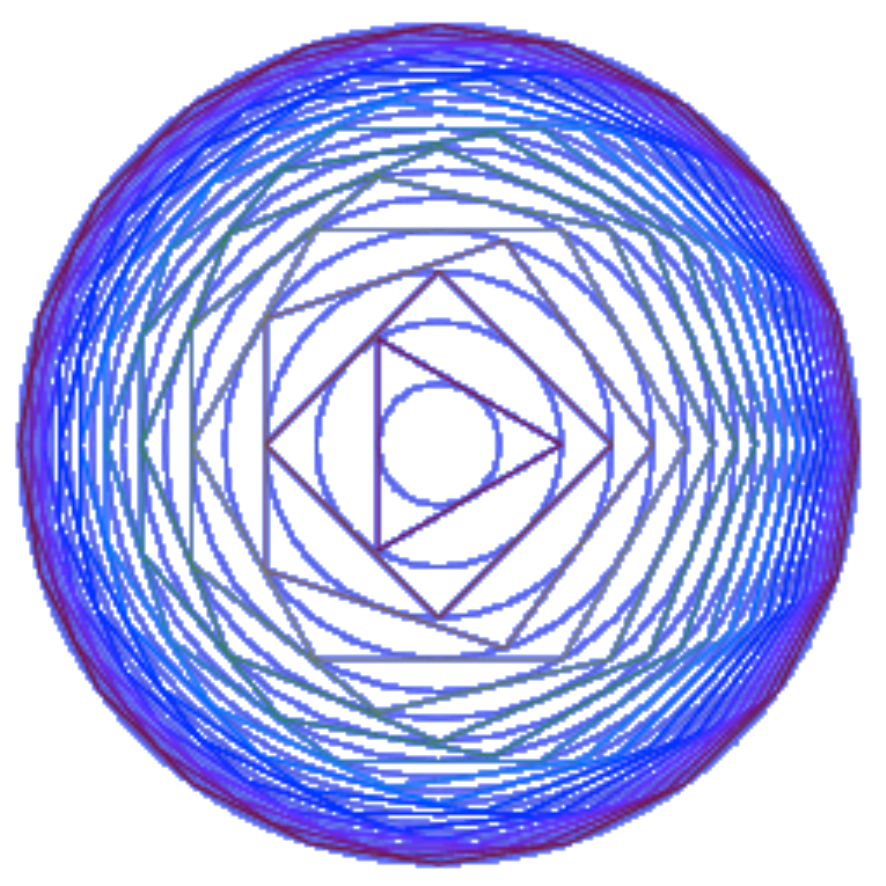}
\caption{(Color online) Result of taking infinitely nested charges at the edges of circumscribed regular polygons. See text for details.}
\label{FigB}
\end{center}
\end{figure}

Can we imagine a system of discrete charged particles --not a continuous one-- which occupies a finite region of space and has 
infinite total charge? The answer is yes. A way to reach infinity without putting more particles in the same place can be found in 
the setting of an interesting problem borrowed from discrete mathematics, that is, nesting circumscribed charged polygons. 
See Fig. (\ref{FigB}). The radius $R_{k}$ of every $k$ shell follows the relation $R_k=\big[\cos\frac{\pi}{k+2}R_{k-1}\big]^{-1}$. Thus, for an initial 
$R_0=1$, the final shell is found at $R_{\infty}=\prod_{k=1}^{\infty} \big[\cos\frac{\pi}{k+2}\big]^{-1}=8.700036.. >1$ (the size of the systems grows 
from the initial value). 

\begin{figure}[htbp]
\begin{center}
\includegraphics[width=8.8cm]{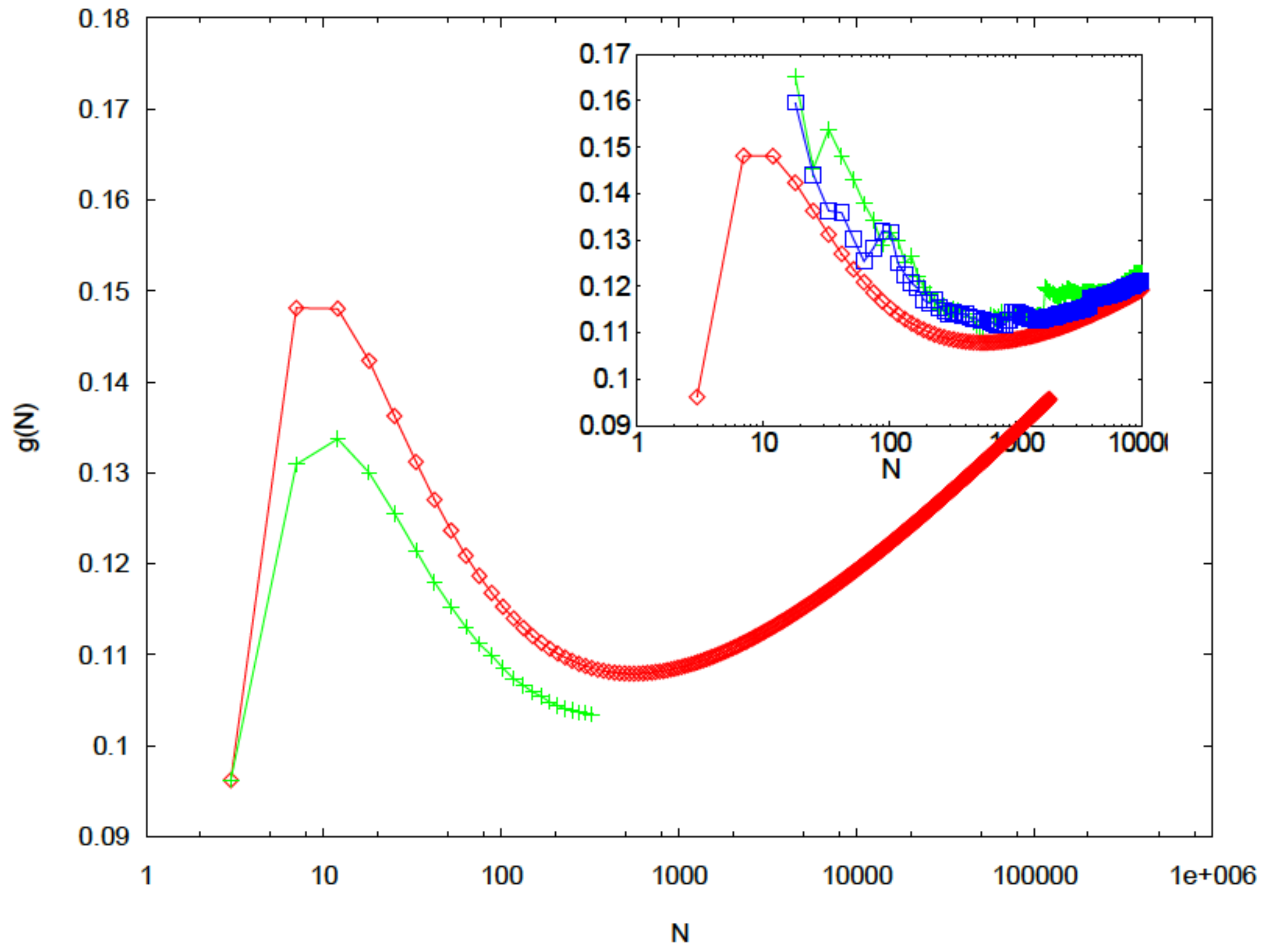}
\caption{(Color online) Plot of the energy in terms of $N^2$ versus $N$. The equilibrium curve has certainly lesser energy that 
the configuration of Fig. (\ref{FigB}), but the behavior is more or less the same. The inset depicts several random shifts of the 
inscribed polygons. An overall tendency, though, is apparent. See text for details.}
\label{FigC}
\end{center}
\end{figure}

\noindent Every shell is filled with $n_k=k+2$ particles at the vertices, while the total number of them up to the $k$-shell is given by 
$N_k=\frac{k(k+5)}{2}$. It can be easily shown that  $R_{k}/R_{\infty} \sim 1 - A/\sqrt{N_k}$, where $A$ is some positive constant. 
The form factor $g(N)=E_N/N^2$ is then computed for $k=490$ shells, which amounts to 99 \% of $R_{\infty}$. In Fig. (\ref{FigC}) 
we depict $g(N)$ versus $N$. It can be appreciated the particular behavior of $E_N/N^2$, having one local maximum and minimum 
(however, $E_N$ is always monotonically increasing with $N$). This particular configuration is not the one reached in equilibrium. 
The lower curve is the one obtained by minimizing over all $2N_k$ positions of the particles. We see that there is not much difference between 
the optimal curve and the setting of the polygons each having one vertex in the positive $x$-axis. It is precisely the shape of $g(N)$ 
makes this system a peculiar one, specially when compared with the previous planar systems. Now, in the inset of Fig. (\ref{FigC}), 
we compare the previously discussed curve with the ones obtained by randomly rotating each polygon with respect to the others. It is quite 
remarkable that a common tendency is followed by all these curves.

The non-monotonic behavior of $g(N)$ that has an intriguing physical meaning since its inverse is twice the capacity $C_N$ of the system. 
This particular evolution can be understood as the following: at first, all particles at the vertices have plenty of space to occupy, but then 
the rate of growth slows down, specially under the effect of an outer background of charges. From there onwards, the energy increases 
at the same rate, which implies that the capacity of the system diminishes progressively.
\vskip 1cm

\begin{figure}[htbp]
\begin{center}
\includegraphics[width=8.8cm]{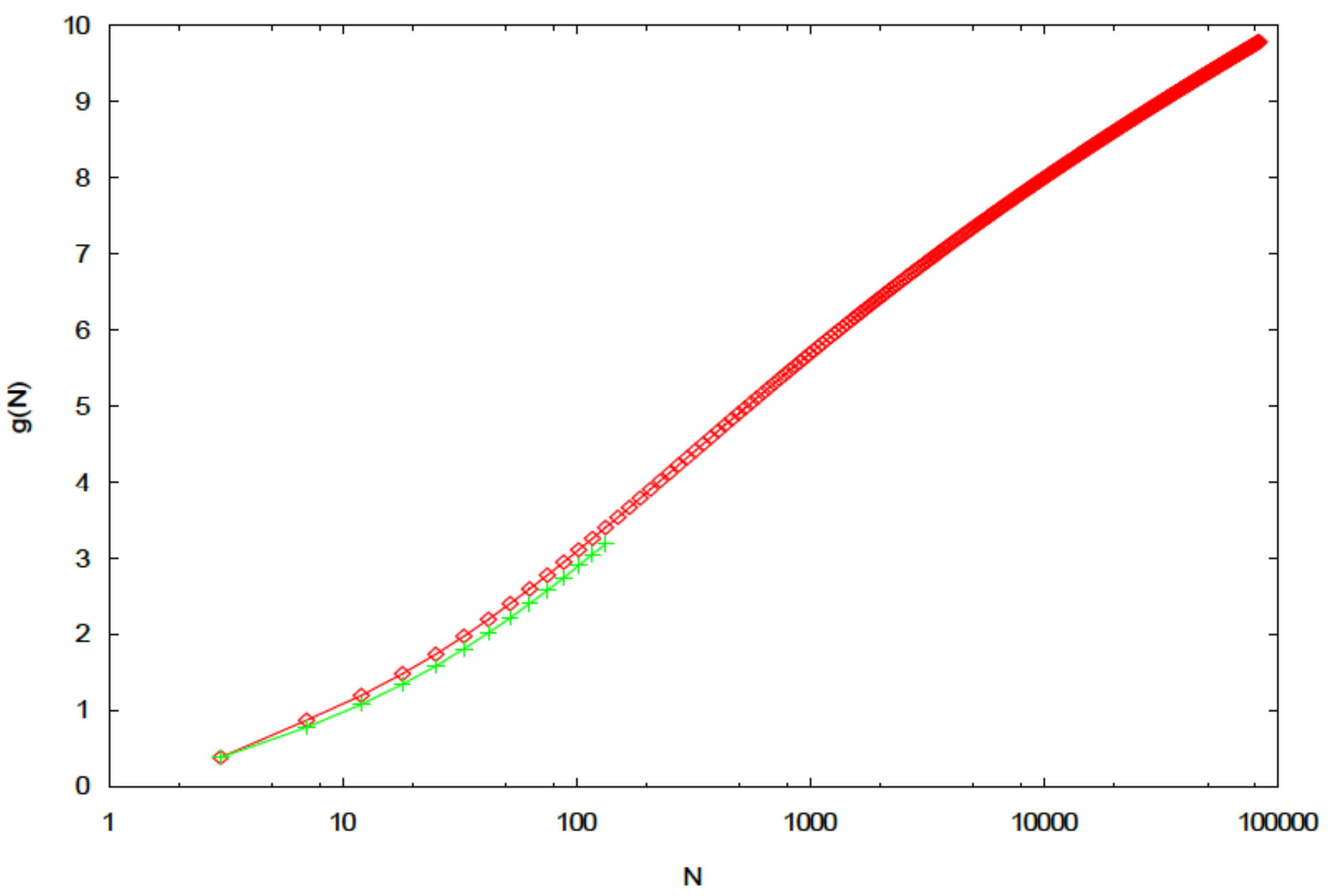}
\caption{(Color online) Growth of the $E_N/N^2$ versus $N$. Notice how the equilibrium curve (lower one) approaches quite 
closely the configuration of inscribed polygons done in a similar fashion as in Fig. (\ref{FigB}). See text for details.}
\label{FigD}
\end{center}
\end{figure}

The instance of inscribed polygons is such that now we have the relation $R_k= \cos \frac{\pi}{k+2}R_{k-1}$. Thus, the final shell ($R_0=1$) 
is reduced to the value $R_{\infty}^{'}=\prod_{k=1}^{\infty} \cos \frac{\pi}{k+2}=1/8.700036.. <1$. In this case, 
$R_{k}/R_{\infty}^{'} \sim 1 + A/\sqrt{N_k}$, where $A$ is the same positive constant encountered before. As opposed to the particular evolution of 
$g(N)$ in Fig. (\ref{FigC}), now we have a very smooth behavior, as seen from Fig. (\ref{FigD}). Also, the energetic equilibrium (lower curve) is quite 
close to the inscription of polygons done in a similar fashion as in Fig. (\ref{FigB}). It is very remarkable that growing inwards, as in the present case, 
differs very much from the situation of growing outwards.

\subsection{Three dimensions}

The three-dimensional euclidean case is indeed the paradigm of the Thomson problem. We take the unit sphere, and then let the particles interact and arrange themselves to that 
the minimize the total repulsive energy with the constriction of being located {\it on the surface} of the $S^2$-sphere. The generalized Thomson problem arises, for example, in determining the arrangements of the protein that may comprise the shells of viruses. The proposed encapsulation of active ingredients such as drugs, made the problem appealing. 

\begin{figure}[htbp]
\begin{center}
\includegraphics[width=8.8cm]{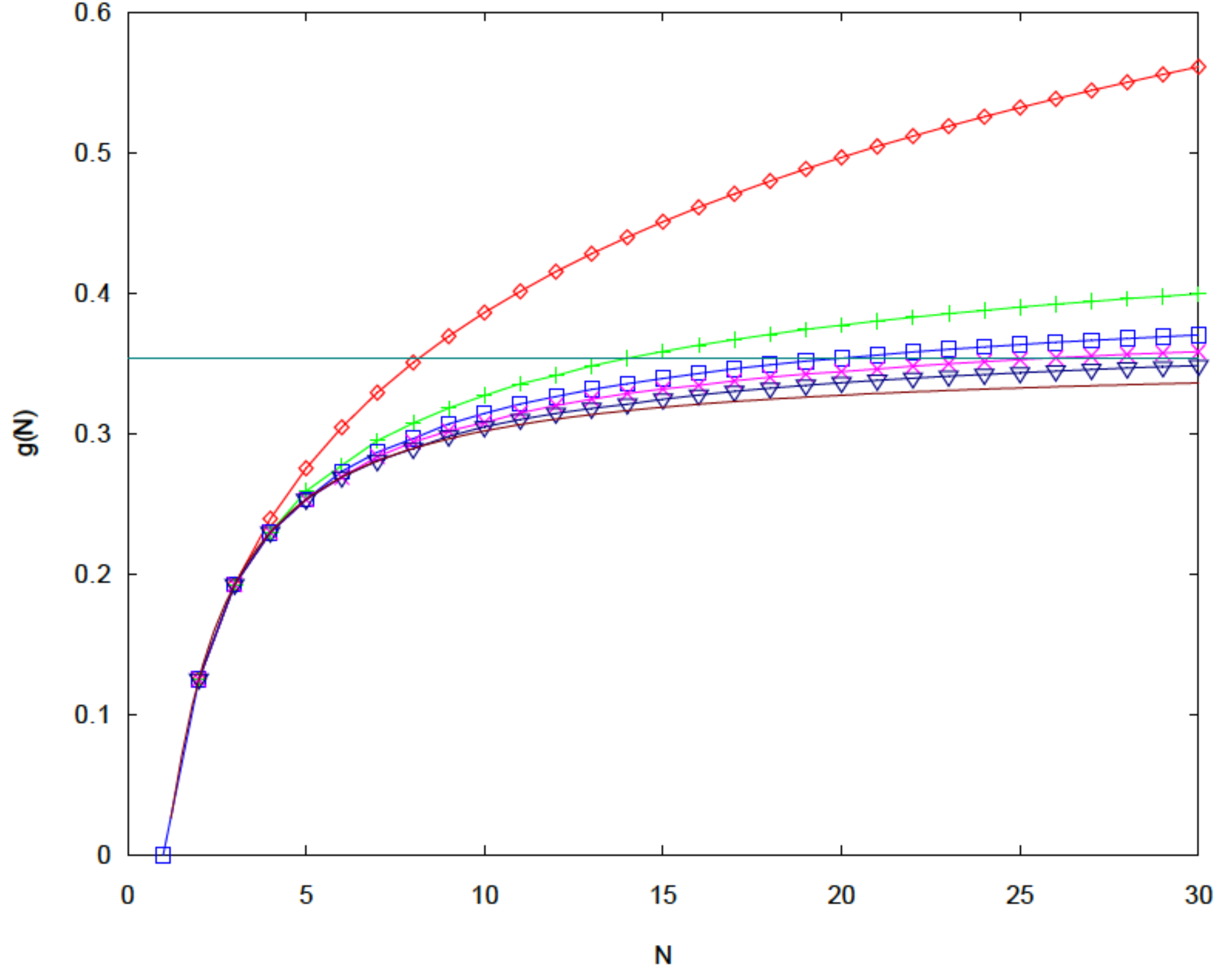}
\caption{(Color online) Growth of the energy in terms of $N^2$ versus $N$ for the $S^2$-sphere (upper curve). The upper curve the $d=2$ one. 
The lower curves correspond to dimensions 
$d=3,4,5,7$ and $\infty$. See text for details.}
\label{FigE}
\end{center}
\end{figure}

\begin{figure}[htbp]
\begin{center}
\includegraphics[width=8.8cm]{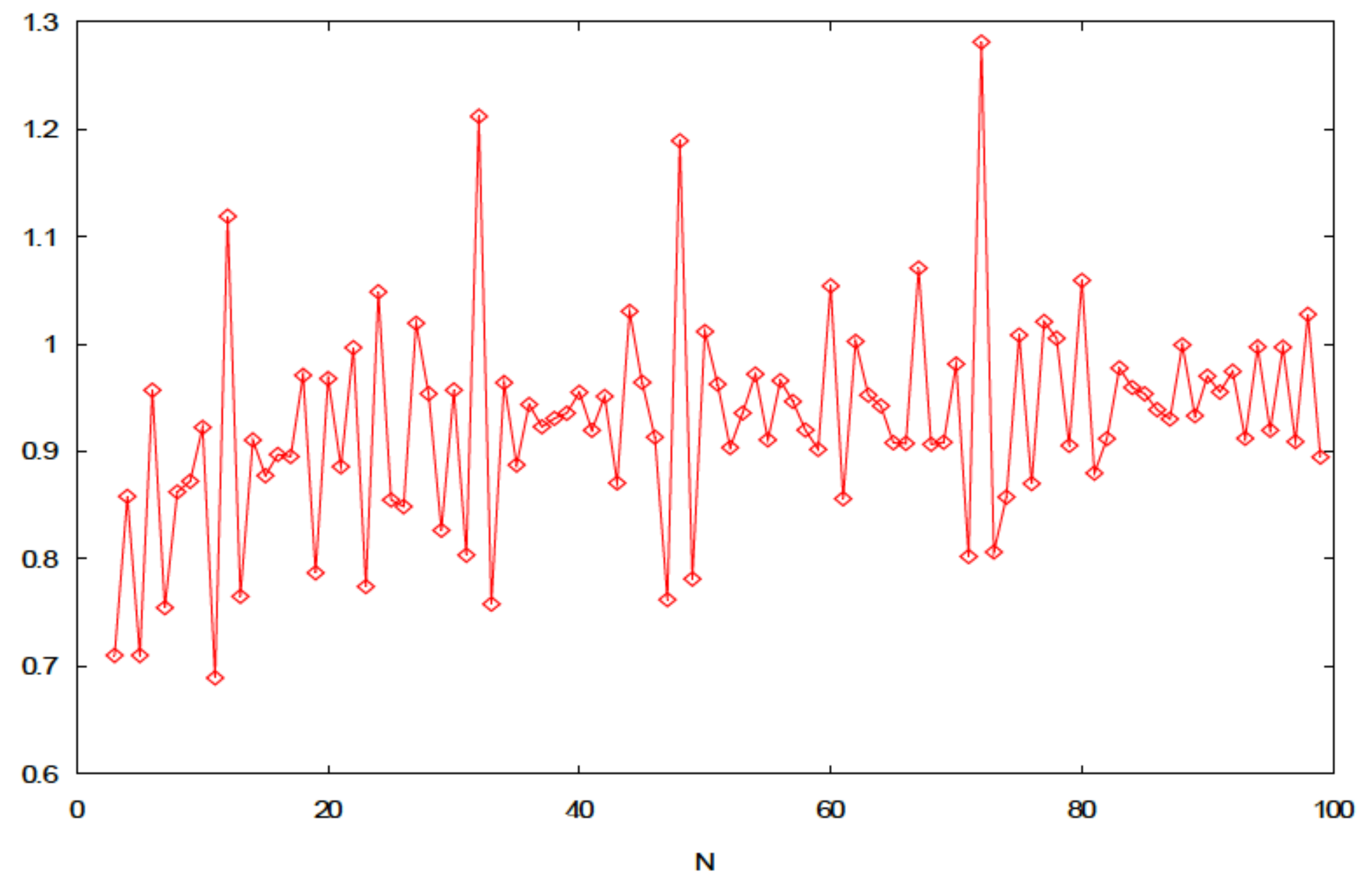}
\caption{(Color online) Plot of $\Delta \mu_N=E_{N+1}+E_{N-1}-2E_N$ versus $N$. Definite peaks arise as being more stable (12-32-48-72). See text for details.}
\label{FigF}
\end{center}
\end{figure}

\noindent The minimization now takes place for the $2N$ degrees of freedom of the charges (bear in mind that they are confined at a fixed distance). The rigorous proof for an arbitrary number 
of constituents is by no means available, so we have to resort to numerical computation. Some small instances, such as the 5-particle case, have been the subject of recent 
study \cite{5electron}. The ensuing results are depicted in Fig. (\ref{FigE}). Notice how smooth is the evolution of energies. In point of fact, the chemical potential 
$\mu_N=E_N-E_{N-1}$ is nearly a straight line and less that $N-1$ $\forall$ $N$. However, when look carefully, we see that $\Delta \mu_N=E_{N+1}+E_{N-1}-2E_N$ displays the most 
intimate structure of the system. As seen from Fig. (\ref{FigF}), there are configurations for which the system is more stable than others. It is plain, then, the peaks unveil configurations that 
are very stable. The tetrahedron (4), octahedron (8) and icosahedron (12) appear to be very robust equilibrium configurations, specially the icosahedron 
(first out of the four prominent peaks seen in the figure 12-32-48-72). These peaks belong to the different symmetry types (point groups in three dimensions). 
$N=12$ and $N=32$ belong to the the chiral icosahedral group, and so does $N=72$. They possess the rotation axes of an icosahedron or dodecahedron and, additionally, $N=12$ and $N=32$ include horizontal mirror planes and contain also inversion center and improper rotation operations. Regarding $N=48$, which belongs to the chiral octahedral group, it has the 
rotation axes of an octahedron or cube. Needless to say, the all have zero dipole values.

\subsection{Four dimensions}

Connection between equilibrium configuration and regular polyhedra has already occurred for the $S^2$-sphere, where we had coincidences with three Platonic solids, 
namely, the tetrahedron, the octahedron and the icosahedron. If hyperspace ($d=4$ in Euclidean space) we have six regular solids, which are called convex polychora \cite{coxeter1,coxeter2}. 
Of the six regular convex polychora, five are typically regarded as being analogous to the Platonic solids: the 4-simplex (a hyper-tetrahedron), the 4-cross polytope 
(a hyper-octahedron), the 4-cube (a hyper-cube), the 600-cell (a hyper-icosahedron), and the 120-cell (a hyper-dodecahedron). The 24-cell, however, has no perfect analogy 
in higher or lower spaces. The pentatope and 24-cell are self-dual, the 16-cell is the dual of the tesseract, and the 600- and 120-cells are dual to each other. 

When minimizing the total energy for different number of particles, we obtain the results depicted in Fig. (\ref{FigE}). Notice the abrupt change in the slope of the curve from 
two dimensions to three dimensions. As in the case of $S^2$, in the $S^3$-sphere we encounter also three convex regular bodies, despite the fact the we have an additional one: 
i) the usual regular simplex or pentatope (5 vertices and 10 edges of $\frac{\sqrt{5}}{2}$ lenght), 
ii) the reciprocal politope of the hypercube --the 16 cel-- (8 vertices and 24 edges of $\sqrt{2}$ length and 
iii) the 600-cell politope (120 vertices and 720 edges of $1/\Phi$ length, $\Phi$ being the Golden Ratio).

\begin{figure}[htbp]
\begin{center}
\includegraphics[width=8.8cm]{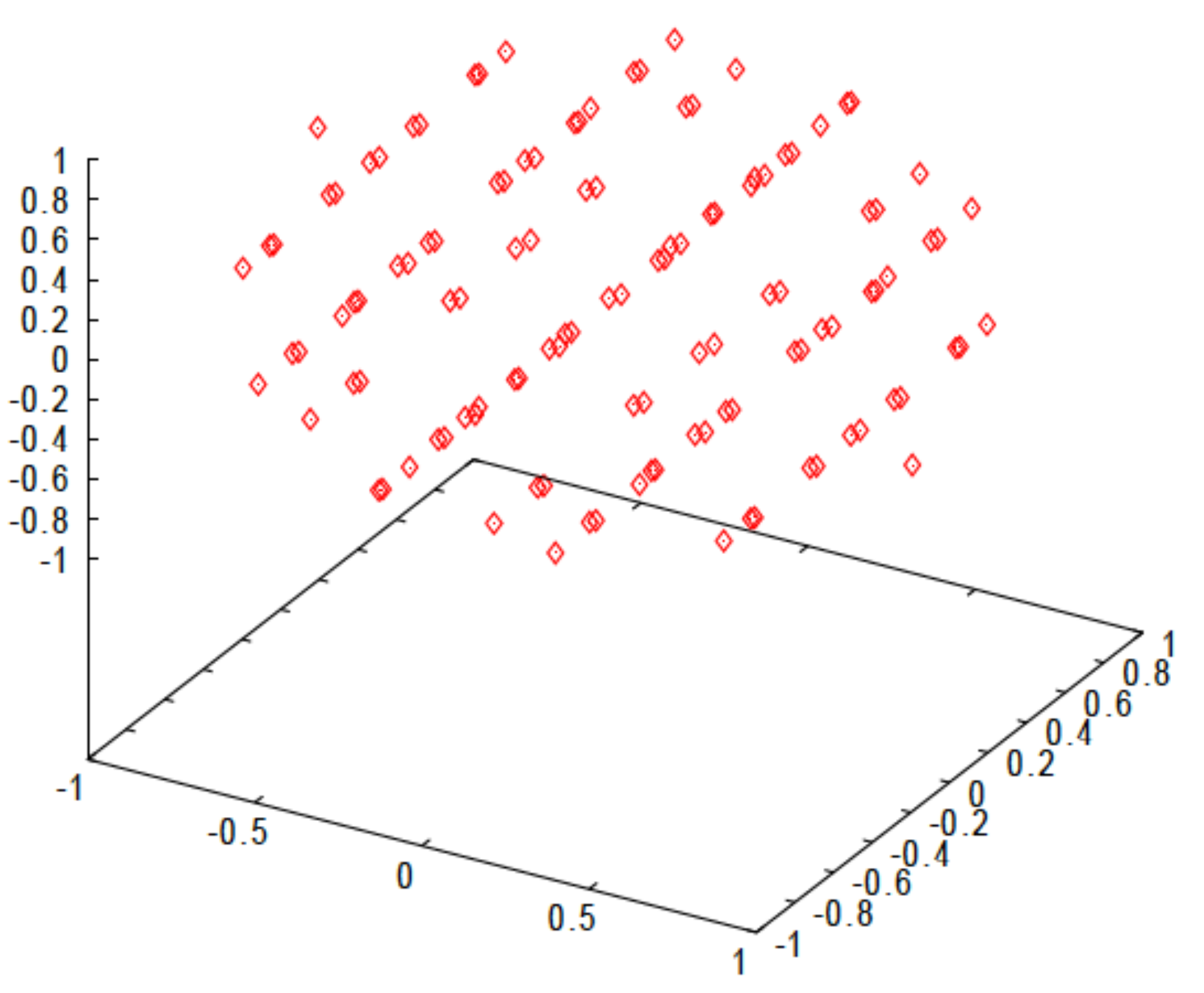}
\caption{(Color online) Plot of the projection onto the x-y-z space of the 600-cell. See text for details.}
\label{FigG}
\end{center}
\end{figure}

\begin{figure}[htbp]
\begin{center}
\includegraphics[width=6.8cm]{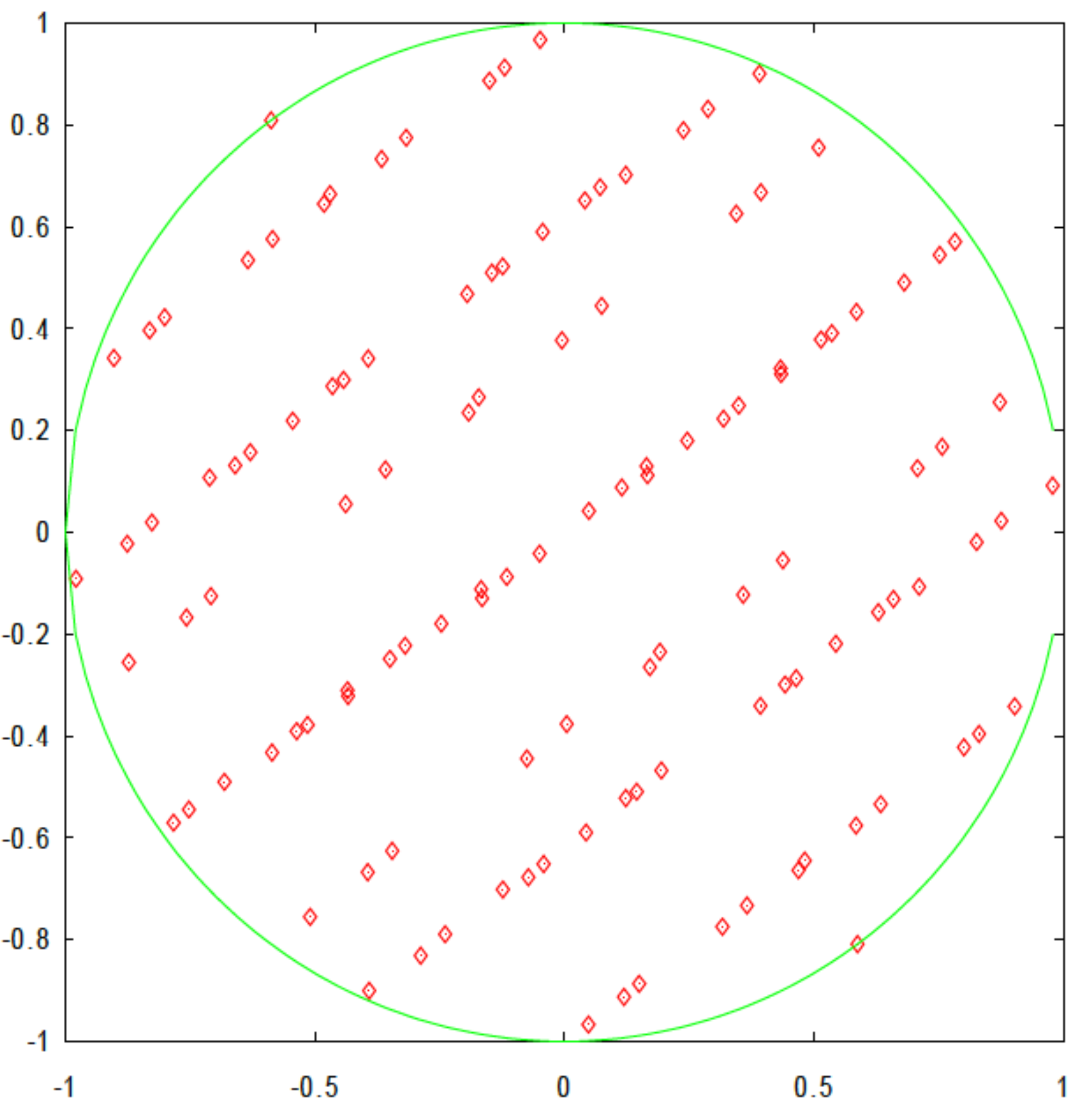}
\caption{(Color online) Plot of the projection onto the x-y plane of the 600-cell. The number of vertices at graph distances from
one vertex to the opposite one (1,12,32,42,31,1) are seen as ``bands''.  See text for details.}
\label{FigH}
\end{center}
\end{figure}

In Fig. (\ref{FigG}) we depict the projection onto the x-y-z 3D space of the 600-cell. Notice the apparent regularity. 
The skeleton of the 600-cell is a 12-regular graph. The number of vertices at graph distances from
one vertex to the opposite one is given by 1,12,32,42,31 and 1, which can be clearly seen as ``bands'' in the projection 
onto the x-y plane that appears in Fig. (\ref{FigH}).

\subsection{Arbitrary dimensions}

Cartesian space in dimensions greater then 4 does not contain those rich structures found for three and four dimensions. The only regular 
bodies that can be found are $d$-simplex forms (extension of the tetrahedron), $d$-cubes and $d$-octahedron. Once we know that, these instances 
might appear in the configurations that minimize the total electrostatic energy of the system in the $S^{d-1}$-sphere. Particularly, the $d$-simplex 
is the only one occurring in the equilibrium configurations.

In Fig. (\ref{FigE}) all curves for $E_N/N^2$ versus $N$ (from top to bottom) are depicted, corresponding to dimensions 
$d=2,3,4,5,7$ and $\infty$. Notice how the slope of the curves flattens as we increase the corresponding dimension of $S^{d-1}$ 
and the number of particles $N$. The coincidences we see for different curves (dimensions) and particle number $N$ are 
explained because one simplex is embedded into the following one ($N=d+1$). Therefore, if we have 
$N$ particles and the length of the simplex is $L=\sqrt{2}\sqrt{\frac{d+1}{d}}$, the ensuing equilibrium energy value between all pairs 
of particles is $E_N=\frac{N(N-1)}{2} \cdot \frac{1}{L}$ or, in a simplified form,

\begin{equation}
\frac{E_N}{N^2}\,=\,\frac{\sqrt{2}}{4}\,\frac{N-1}{N}\,\sqrt{\frac{N}{N+1}}.
\end{equation}

\noindent The horizontal line in Fig. (\ref{FigE}) corresponds to the limiting value $\frac{\sqrt{2}}{4}=0.35355..$

Notice that the evolution of the curves in Fig. (\ref{FigE}) is such that they saturate very fast until $d=7$, 
which is very close to the case of
infinite dimensions. Whether this issue is related or not to the fact that the surface of the $S^{d-1}$-sphere
peaks around $d=7$ and steadily goes to 0 for greater values is something that has to be
carefully analyzed. From our results, one can wonder how come the surface reduces yet there is
enough space for the energy to diverge. The only possible explanation is that since the distances
between particles tend to a fixed value ($\sqrt{2}$), the surface should not decrease faster that the
allocation of the particles inside. We should compare how the surface diverges for $d$ large and then
compare the surface energy with the term $\frac{\sqrt{2}}{4}\cdot N^2$ we have. 

\noindent In any case, going to greater and greater dimensions imply that particles arrange themselves inside very packed 
structures. The greater the dimension and the number of particles, the more constant is the capacity $C_N$, which is related to 
the inverse of $E_N/N^2$, which constitutes a very surprising result.

\section{The Thomson problem for several geometric confinements}

\subsection{The Thomson problem on the cube}

Suppose that we allow particles to stay on the surface of a cube of length 2, so that an sphere of unit radius can be perfectly inscribed. 
The study of the corresponding problem of how particles reach an equilibrium under the sole interaction of a Coulombian repulsive 
force will certainly lead to structures with some symmetry (irregular polyhedra, in fact), but very different to the ones encountered in the 
$S^2$-sphere. 

\begin{figure}[htbp]
\begin{center}
\includegraphics[width=8.8cm]{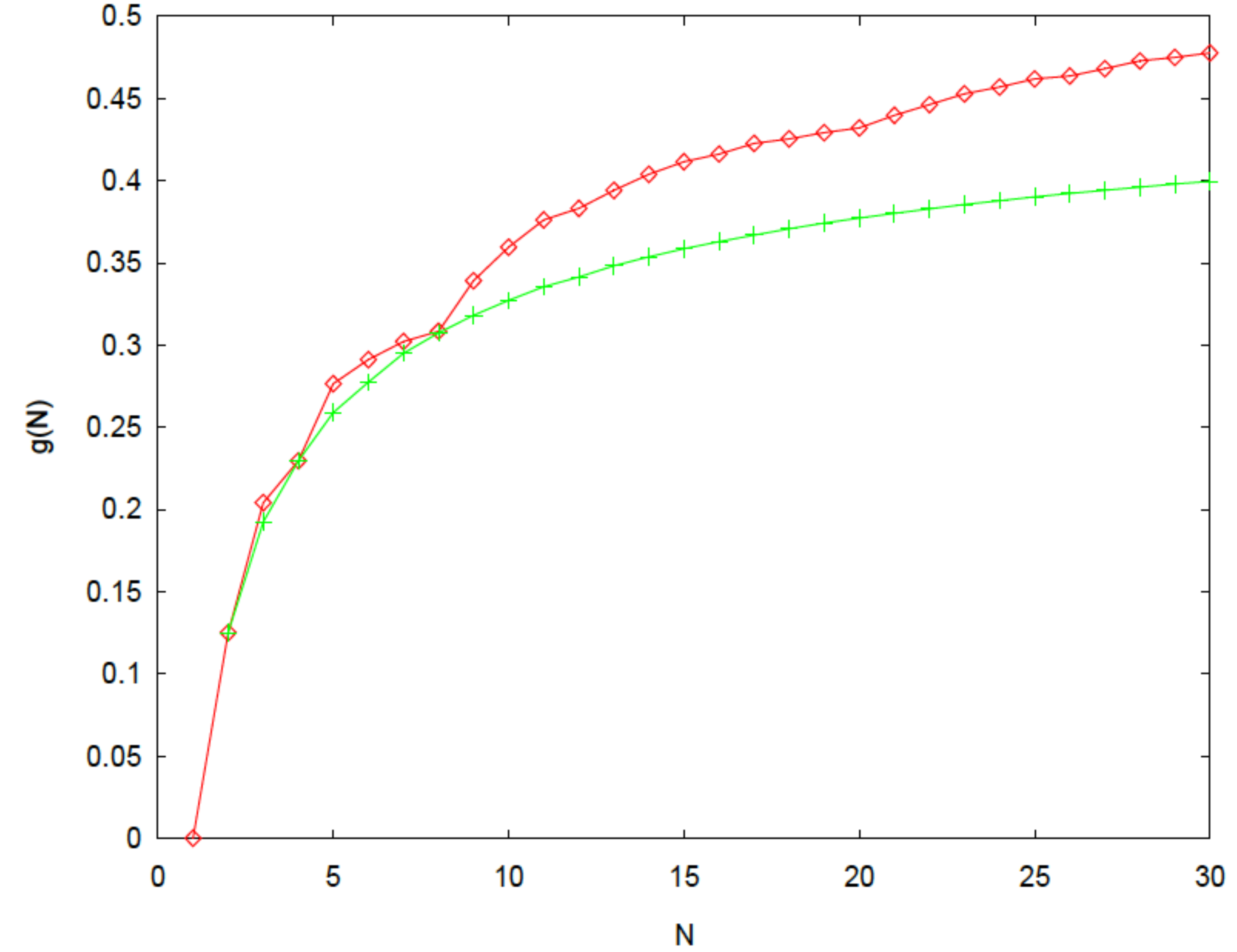}
\caption{(Color online) Plot of $E_N/N^2$ versus $N$ for the cubical confinement. The lower curve corresponds to the $S^2$-sphere. 
The cube dimensions have been modified to order to be embedded inside the unit sphere. See text for details.}
\label{FigI}
\end{center}
\end{figure}

In Fig. (\ref{FigI}) the ensuing values of $E_N/N^2$ are depicted versus $N$, as well as those of the embedded unit sphere. 
As expected, we only have three coincidences. The evolution of the curve is such that a clear shell structure appearing on the cube 
can be inferred. The general analytical result is far from being reachable. However, it appears that some periodic tendency occurs, as seen 
from the dips for several $N$-values. A related study will be found elsewhere \cite{noltrosfeina}.

\subsection{The Thomson problem in ``jelly-like'' charged systems}

The problem of letting charges to spread inside the $S^2$-sphere and not only on the boundary does not show any energetically
advantage here, at least for the systems studied up to $N=200$ particles. When considering the same number of 
particles $N$, two different configurations are considered: one with all charges on the surface (usual Thomson problem) and another 
one with one particle at the origin, surrounded by the $N-1$ remaining charges. The system will choose which one of these two possible 
configurations is better because of an energetic argument. Due to spherical symmetry $E_{N}^{*}$ (origin)$=E_{N-1} + (N-1)\cdot \frac{1}{R}$, 
$E_{N-1}$ being the equilibrium energy for $N-1$ particles in the usual Thomson problem. 

\begin{figure}[htbp]
\begin{center}
\includegraphics[width=9.0cm]{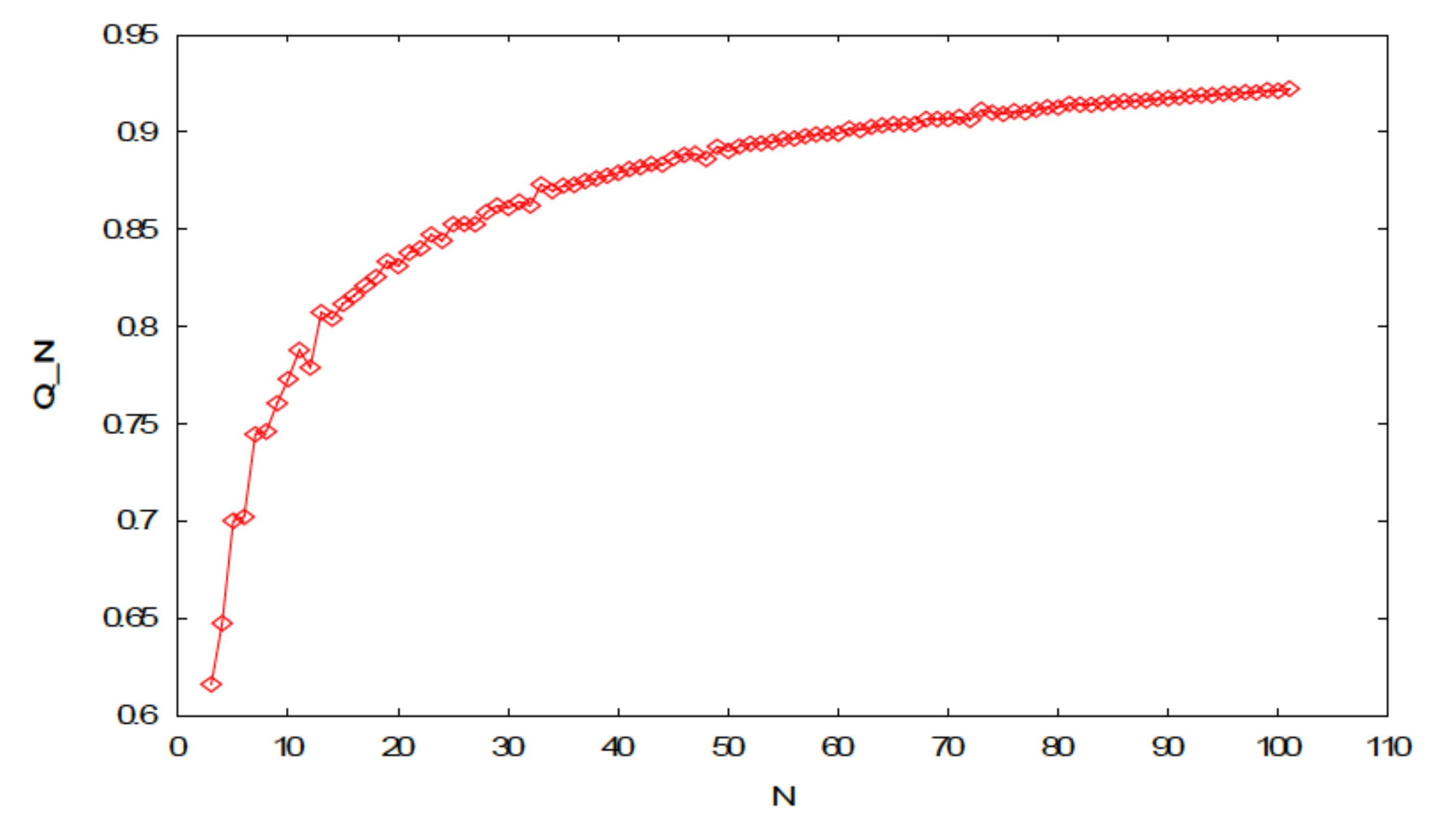}
\caption{(Color online) Plot of the quantity $Q_N$. See text for details.}
\label{FigJ}
\end{center}
\end{figure}

To know wether the system starts to allow particles inside (the center, to be more precise) the
quantity $E_{N}^{*} - E_{N}$ must be negative, which is tantamount as saying that
$Q_N \equiv ( E_{N} - E_{N-1} ) / ( N - 1 ) > 1$. In other words, the chemical potential $\mu_N$ for the original
problem should at some point be greater than $N-1$. The quantity $Q_N$ is depicted in Fig. (\ref{FigJ}).  
As we can appreciate, the tendency of the curve is to remain below 1. But not only this, we also obtain from
the picture information regarding stability: some $N$-configurations are more stable than others.

Thus, surprisingly, the jelly-like system of charges is not more stable than the one with all particles
lying on the surface. The case where the number of particles is greater than 200 has not been
studied, but it is plausible that for large $N$, some nested configuration of charges might be
energetically favored, but very unlikely to occur.

The previous result has been numerically checked for many different configurations allowing more
than one charge inside. Interestingly enough, results obtained in 2D \cite{noltrosfeina} show the system is more stable 
having charges with $r < R$. There is, then, no jelly-like possible configurations for $d > 2$! 
As a matter of fact, the unexplained change in the slope from $d=2$ to $d>2$ in Fig. (\ref{FigE}) 
and the similarity of curves among greater
dimensions might be the reason why no inner structure is allowed there. This is a really surprising result. In
fact, in $d=2$ the system acts as a metal. To be more precise, the energy differences per particle between
configurations with all charges on the surface and those ones with one or more particles inside tend
to differ less and less as $N$ increases. That is, slight perturbations may allow particles to scape the
surface and enter the inner region almost in a continuous way. The difference between the 2D
dimension and the rest is that the first one do possess less energy for configurations with particles
inside, whereas this is not the case for $d > 2$ though energies between equilibrium configurations and
``excited states'' stay quite close.

\section{The Thomson problem in non-euclidean geometries}

The mathematical works regarding multidimensional geometries and non-Euclidean geometries had been considered over 
the years of their inception are mere mathematical abstractions until Poincar\'e proved that the group of Lorentz transformations 
that rendered Maxwell's equations invariant could be regarded as rotations in an space of four dimensions. Later on, the works 
of Einstein and the corresponding geometrical insight by Minkowski led to accept the fourth dimension as a necessary description 
for those phenomena related to electromagnetism. 

In mathematics, non-Euclidean geometry is a small set of geometries based on axioms closely related to those specifying Euclidean geometry. 
As Euclidean geometry lies at the intersection of metric geometry and affine geometry, non-Euclidean geometry arises when 
either the metric requirement is relaxed, or the parallel postulate is set aside. In the latter case one obtains hyperbolic geometry and elliptic geometry, 
the traditional non-Euclidean geometries. When the metric requirement is relaxed, then there are affine planes associated with the planar 
algebras which give rise to kinematic geometries that have also been called non-Euclidean geometry.

In the present Section we shall study how the choice of the geometry where the system of particles is embedded affects the ensuing 
equilibrium configurations under the Coulombian interaction.

\begin{figure}[htbp]
\begin{center}
\includegraphics[width=8.6cm]{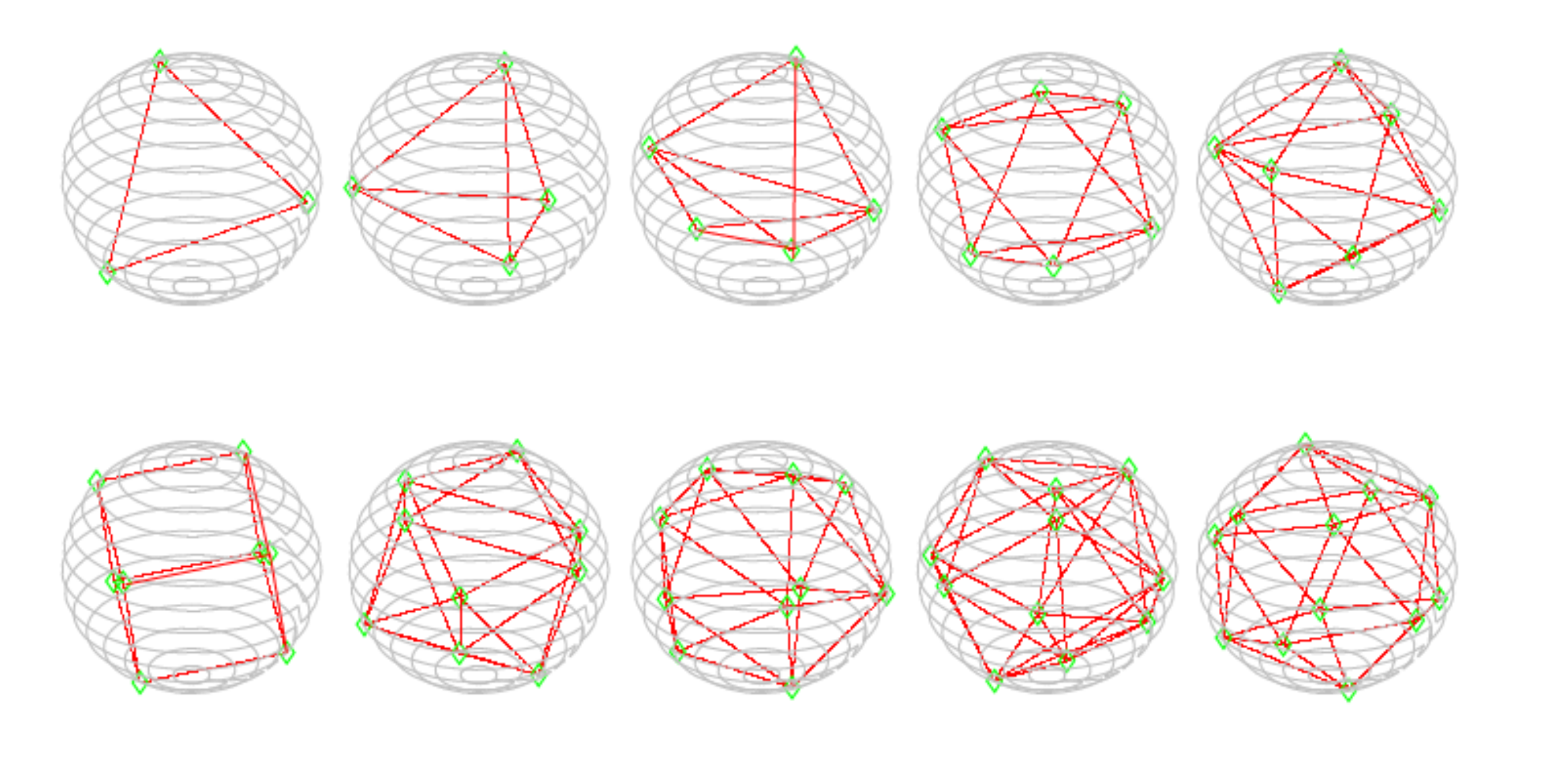}
\caption{(Color online) Series of plots depicting, from left to right and top to bottom, the cases $N=3..12$ in the elliptic plane. 
The red lines show the structure of the regular polygons that appear on the surface once these lines are projected on it. Almost all 
structures are shared with the ones that appear in the usual Thomson problem in the euclidean space, except for $N=8$, where in this 
case we have a regular polyhedron (a cube), contrary to the 3D euclidean case (regular square antiprism). See text for details.}
\label{FigX}
\end{center}
\end{figure}

\subsection{The elliptic plane and space}

The elliptic plane is modeled by the $S^2-$sphere, where distances between points are taken, in the unit sphere, as the angles between points 
measured along geodesics (great circles in this case). The ensuing optimization over all particles provide very interesting results, as seen 
in Fig. (\ref{FigX}) . For $N=2$, we obtain two antipodal points, whereas for $N=3$ we have the sphere cut into two halves. For $N=4$ we have a tetrahedron, 
which gives rise to 4 equal spherical triangles. For $N=5$ we obtain a triangular bipyramid, consisting of 6 equal spherical triangles. In the case 
of $N=6$, we naturally obtain an octahedron (8 equal spherical triangles).
The $N=7$ instance provides us with regular pentagonal bipyramid (10 spherical triangles). The $N=8$ case marks a big difference with regards 
the 3D euclidean case, where it does not occur. In this case, 6 regular spherical squares cover the whole surface. $N=9$ has a structure of a 
triaugmented triangular prism, from by 14 faces from two different spherical triangles. $N=10$ provides us with a gyroelongated square bipyramid structure, 
consisting of 16 regular spherical triangles. The last figure studied which bears some relevances is the $N=12$ case, the icosahedron.

On the whole, only three out of the five platonic solids appear in the 3D euclidean case, whereas two more does in the elliptic plane. Although not shown here, 
$N=12$ particles arranges themselves forming a dodecahedron, thus confirming the appearance of all five platonic bodies in the minimization of the total coulombian 
repulsion energy between particles on the elliptic plane. 
\vskip 1cm

\begin{figure}[htbp]
\begin{center}
\includegraphics[width=9.0cm]{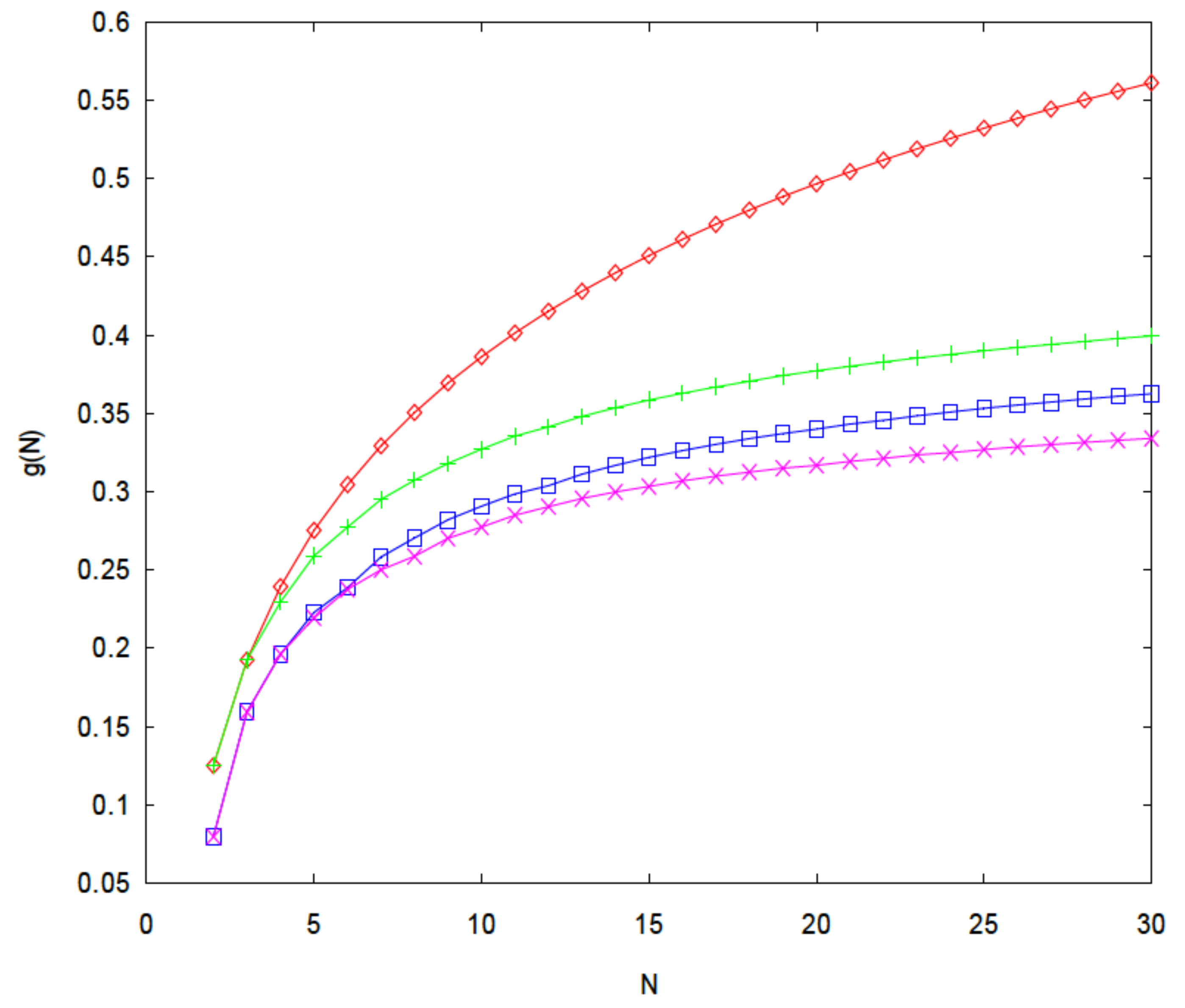}
\caption{(Color online) Dependence of $g(N)=E_N/N^2$ vs. $N$ for the elliptic plane and space (two lower curves, respectively). As we increase the 
dimension, the tendency is to decrease to total energy factor $g(N)$, in the same fashion as in the euclidean case (plane and space, two curves above). 
See text for details.}
\label{FigPC}
\end{center}
\end{figure}

Now, the elliptic space is modeled by the $S^3-$sphere, where distances between lines are taken in the unit hypersphere as angles between lines (great circles). 
The comparison between the elliptic plane and space is shown in Fig. (\ref{FigPC}). $E_N/N^2$ is depicted versus $N$ for the elliptic plane and space (lower curves), 
as compared to corresponding curves in the euclidean plane and space (upper curves). Is it apparent that the tendency of $E_N/N^2$ to diminish with $N$ as we increase the dimension is not particular of the euclidean metric. Furthermore, the 3D euclidean and elliptic space share a similar asymptotic behavior and 
$N$ increases. This particular instance remains unexplained.

\begin{figure}[htbp]
\begin{center}
\includegraphics[width=4.8cm]{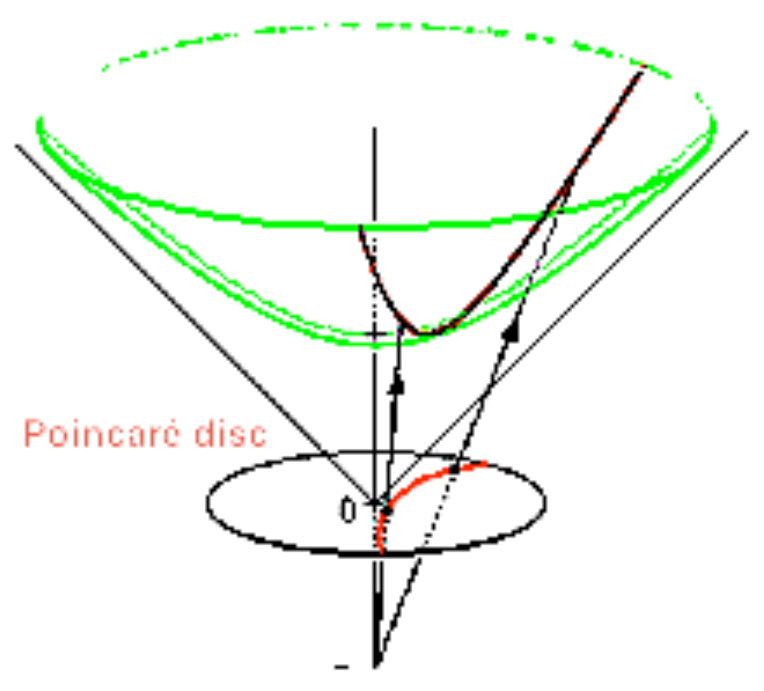}
\caption{(Color online) The hyperbolic model. See text for details.}
\label{FigHyper}
\end{center}
\end{figure}

\begin{table}[h]
\begin{center}
\begin{tabular}{|c||c|c|c||c|c|}
  \hline
  $N$ & $Shells$ & $N$ & $Shells$ & $N$ & $Shells$ \\
  \hline
  1 & 1 &  11 & 2,\,9 & 21 & 5,\,16\\
  2 & 2 &  12 & 2,\,10 & 22 & 1,\,6,\,15\\
  3 & 3 &  13 & 2,\,11 & 23 & 1,\,6,\,16\\
  4 & 4 &  14 & 3,\,11 & 24 & 1,\,6,\,17\\
  5 & 5 &  15 & 3,\,12 & 25 & 1,\,7,\,17\\
  6 & 1,\,5 &  16 & 4,\,12 & 26 & 1,\,7,\,18\\
  7 & 1,\,6 &  17 & 4,\,13 & 27 & 1,\,8,\,18\\
  8 & 1,\,7  &  18 & 5,\,13 & 28 & 1,\,9,\,18\\
  9 & 1,\,8  &  19 & 5,\,14 & 29 & 1,\,8,\,20\\
  10 & 1,\,9  &  20 & 5,\,15 & 30 & 1,\,8,\,21\\
   \hline
\end{tabular}
\end{center}
\caption{Mendeleev table for the harmonic oscillator confinement in the Hyperbolic space of interacting charged particles. Configurations 
result as projections of the hyperboloid model onto the Poincar\'e disc. Results greatly differ from those in the Euclidean plane or space. 
See text for details.} 
\label{taula}
\end{table}

\subsection{The hyperbolic space}

Models have been constructed within Euclidean geometry that obey the axioms of hyperbolic geometry, thus proving that the parallel 
postulate is independent of the other postulates of Euclid (assuming that those other postulates are in fact consistent). 
We shall use the hyperboloid model, also known as the Minkowski model or the Lorentz model, is a model of $n$-dimensional 
hyperbolic geometry in which points are represented by the points on the forward sheet a two-sheeted hyperboloid 
in $(n+1)$-dimensional Minkowski space. The model is depicted in Fig. (\ref{FigHyper}), where we see the connection between 
the hyperboloid and a planar circle called Poincar\'e disc, which is an equivalent representation. 

The metric used in the hyperbolic model is the following one: $z^2-x^2-y^2=1$. This space has negative Gaussian curvature 
$K(z)\,=\,-\frac{1}{(1-2z^2)^2}$. The corresponding parameterization is given by

\begin{eqnarray} \label{hyper}
 x &=& \sinh \mu\,\cosh \nu \cr
 y &=& \sinh \mu\,\sinh \nu\cr
 z &=& \cosh \mu,
\end{eqnarray}

\noindent with $\mu \in (-\infty,\infty)$, $\nu \in [0,\pi)$. The corresponding mapping into Poincar\'e disc is made using $\big(x'=\frac{x}{1+z},y'=\frac{y}{1+z}\big)$. 
Distances between points $P(x_1,y_1,z_1)$ and $Q(x_2,y_2,z_2)$ are given by $d(P,Q)=\cosh^{-1}[z_1z_2-x_1x_2-y_1y_2]$. 
With these tools, we can now study how charged particles arrange themselves under electrostatic interaction. However, since space is not 
limited to a finite region, in the present case we must introduce some overall harmonic oscillator interaction to prevent particles from spreading indefinitely. Otherwise, 
all particle should remain at the infinite or, equivalently,  on the circumference of the Poincar\'e disc in Fig. (\ref{FigHyper}).
Results are shown in Table (\ref{taula}). As we can appreciate, these result differ quite a lot from the ones existing in the literature. Remarkably, 
these results greatly depart from either Euclidean planar of space harmonic oscillator confinements. Therefore, and quite surprisingly, 
the action of the Coulomb interaction in the hyperbolic model results in very different outcomes as compared with other geometries.

\section{Conclusions}

The Thomson problem has been thoroughly revisited here in the light of the growing interest in non-interacting systems in either 2D and 3D dimensions. 
The intimate connection between physics and geometric symmetry has been explained not only in the usual euclidean space. We have shed new light on 
how to describe the Thomson problem in different non-euclidean geometries. An interesting result has been reached for $d \rightarrow \infty$ in the usual 
$\mathbb{R}^d$ space, where particles tend to arrange themselves in extremely packed structures.

\section*{Acknowledgements}

J. Batle acknowledges partial support from the Physics Department, UIB. J. Batle acknowledges fruitful discussions 
with J. Rossell\'o, Maria del Mar Batle and Regina Batle.


\begin{thebibliography}{}


\bibitem{Wig} E.P. Wigner, Phys. Rev. B {\bf 46}, 1002 (1934).

\bibitem{Grimes} C. C. Grimes and G. Adams, Phys. Rev. Lett. {\bf 42}, 795 (1979).

\bibitem{Andrei} E. Y. Andrei, G. Deville, D. C. Glattli, F. I. B. Williams, E. Paris, and B. Etienne, Phys. Rev. Lett. {\bf 60}, 2765 (1988).

\bibitem{Deshpande} V. V. Deshpande and M. Bockrath, Nature Physics {\bf 4}, 314 (2008).

\bibitem{tom} J. J. Thomson, Phil. Mag. {\bf 7}, 237 (1904).

\bibitem{loz}
             Yu.E. Lozovik and V.A. Mandelshtam,
	     Phys. Lett. A{\bf 165}, 469 (1992).
\bibitem{bol}
             F. Bolton and U. R\"ossler,
	     Superlatt. Microstruct. {\bf 13}, 139 (1992).

\bibitem{pet1}
             V. Bedanov and F.M. Peeters,
             Phys. Rev. B {\bf 49}, 2667 (1994).
             
\bibitem{pet2}
             M. Kong, B. Partoens, A. Matulis, and F.M. Peeters,
             Phys. Rev. E {\bf 69}, 036412 (2004).

\bibitem{kirkpatrick83}
Kirkpatrick S, Gelatt Jr C D and Vecchi M P 1983
Science {\bf 220} (4598), 671.

\bibitem{Szego}  G. Szeg\"o, {\it Orthogonal Polynomials}, Amer. Math. Soc. Colloquium Publications, 1975.

\bibitem{Dehesa} R. \'Alvarez-Nodarse and J. S. Dehesa, Applied Mathematics and Computation {\bf 128}, 167 (2002).

\bibitem{Ismail} Mourad E. H. Ismail, Pacific Journal of Mathematics {\bf 193}, 356 (2000).


\bibitem{5electron} 
R. E. Schwartz, {\it Experimental Mathematics} {\bf 22}, 157 (2013).

\bibitem{coxeter1} H. S. M. Coxeter, {\it Math. Z.} {\bf 46}, 380 (1940).

\bibitem{coxeter2}  H. S. M. Coxeter 1969 {\it Introduction to Geometry} (New York: Wiley) 

\bibitem{noltrosfeina} J. Batle et al. Under preparation (2014).



\end{thebibliography}
\end{document}